\input harvmac
\newcount\figno
\figno=0

\input epsf

\def\fig#1#2#3{
\par\begingroup\parindent=0pt\leftskip=1cm\rightskip=1cm\parindent=0pt
\baselineskip=12pt
\global\advance\figno by 1
\midinsert
\epsfxsize=#3
\centerline{\epsfbox{#2}}
\vskip 13pt
{\tenpoint \it Fig. \the\figno: #1}\par
\endinsert\endgroup\par
}

\def\FIG#1#2#3#4{
\par\begingroup\parindent=0pt\leftskip=1cm\rightskip=1cm\parindent=0pt
\baselineskip=13pt
\global\advance\figno by 1
\midinsert
\epsfxsize=#3
\centerline{\epsfbox{#2}}
\vskip #4
{\tenpoint \it Fig. \the\figno: #1}\par
\endinsert\endgroup\par
}

\def\figlabel#1{\xdef#1{\the\figno}}
\def\encadremath#1{\vbox{\hrule\hbox{\vrule\kern8pt\vbox{\kern8pt
\hbox{$\displaystyle #1$}\kern8pt}
\kern8pt\vrule}\hrule}}

\input tables
\thicksize=0.7pt
\thinsize=0.5pt

\def\mystrut{\vrule height 3.7ex depth 1.6ex width 0pt}
\def\mb{\kern-1.2em}
\def\bb{\kern-1.34em}
\def\hb#1{\hbox{#1}}

\def\newline{\hfill\break}
\def\mycomm#1{\hfill\break
$\phantom{a}$\kern-3.5em{\tt===$>$ \bf #1}\hfill\break}
\def\naive{na\"{\i}ve}
\def\wQ{\hbox{(w.~Q)}}
\def\woQ{\hbox{(w/o~Q)}}

\catcode`\@=11
\let\rel@x=\relax
\newcount\timecount
\newcount\hours \newcount\minutes  \newcount\temp \newcount\pmhours
\hours = \time
\divide\hours by 60
\temp = \hours
\multiply\temp by 60
\minutes = \time
\advance\minutes by -\temp
\def\hour{\the\hours}
\def\minute{\ifnum\minutes<10 0\the\minutes
            \else\the\minutes\fi}
\def\clock{
\ifnum\hours=0 12:\minute\ AM
\else\ifnum\hours<12 \hour:\minute\ AM
       \else\ifnum\hours=12 12:\minute\ PM
            \else\ifnum\hours>12
                 \pmhours=\hours
                 \advance\pmhours by -12
                 \the\pmhours:\minute\ PM
                 \fi
            \fi
         \fi
\fi
}

\def\monthname{\rel@x\ifcase\month 0/\or January\or February\or
   March\or April\or May\or June\or July\or August\or September\or
   October\or November\or December\else\number\month/\fi}

\def\bold#1{\setbox0=\hbox{$#1$}     \kern-.025em\copy0\kern-\wd0
     \kern.05em\copy0\kern-\wd0
     \kern-.025em\raise.0433em\box0 }

\def\lsim{\mathrel{\mathpalette\@versim<}}
\def\gsim{\mathrel{\mathpalette\@versim>}}
\def\@versim#1#2{\vcenter{\offinterlineskip
        \ialign{$\m@th#1\hfil##\hfil$\crcr#2\crcr\sim\crcr } }}
\catcode`\@=12
\def\lf{16\pi^2}

\def\frak#1#2{{\textstyle{{#1}\over{#2}}}}
\def\frakk#1#2{{{#1}\over{#2}}}

\def \in{\leftskip = 40 pt\rightskip = 40pt}

\def \out{\leftskip = 0 pt\rightskip = 0pt}

\def\ga{\gamma}

\def\sic{supersymmetric}

\def\NSVZ{{\rm NSVZ}}
\def\DRED{{\rm DRED}}

\def\cmp{Comm.\ Math.\ Phys.\ }

\def\npb{{Nucl.\ Phys.\ }{\bf B}}

\def\plb{{Phys.\ Lett.\ }{\bf B}}

\def\sjnp{Sov.\ J.\ Nucl.\ Phys.\ }

\def\boxit#1{\hbox{\lower0.3em\vbox{\hrule\hbox{\vrule\kern0.15em{\kern0.15em
\lower0.6em\hbox{#1}\vrule height0.3em depth 0.9em width 0pt
\kern0.15em}\kern0.15em\vrule}\hrule}}}
{\nopagenumbers
\vbox{\baselineskip=11pt
\line{\hfil CERN-TH/97-267}
\line{\hfil LTH 401}
\line{\hfil TAUP 2462-97.B}
\line{\hfil hep-ph/9710302}
}
\vskip .3in
\centerline{\titlefont Asymptotic Pad\'e Approximant Predictions:}
\vskip0.1cm
\centerline{\titlefont up to Five Loops in QCD and SQCD}
\vskip 0.3in
\centerline{\bf J.~Ellis}
\medskip
\centerline{\it Theory Division, CERN, CH-1211 Geneva 23, Switzerland}
\centerline{e-mail: \tt John.Ellis@cern.ch}
\bigskip
\centerline{\bf I.~Jack, D.R.T.~Jones }
\medskip
\centerline{\it Dept. of Mathematical Sciences,
University of Liverpool, Liverpool L69 3BX, UK}
\centerline{e-mail:
{\tt dij@amtp.liv.ac.uk} \ and \ {\tt drtj@amtp.liv.ac.uk}}
\bigskip
\centerline{\bf M.~Karliner}
\medskip
\centerline{\it  Raymond and Beverly Sackler Faculty of Exact Sciences}
\centerline{\it School of Physics and Astronomy, Tel-Aviv University,
69978 Tel-Aviv, Israel}
\centerline{e-mail: \tt marek@vm.tau.ac.il}
\bigskip
\centerline{\boxit{\bf M.A.~Samuel}$\,$\foot{Deceased}}
\medskip
\centerline{\it Dept. of Physics, Oklahoma State University,
Stillwater, Oklahoma 74078, USA}
\centerline{\it and}
\centerline{\it SLAC, Stanford University, Stanford, California 94309, USA}

\vskip .2in

\vbox{\baselineskip=13pt
We use Asymptotic Pad\'e Approximants (APAP's) to predict the
four- and five-loop $\beta$
functions
in QCD and $N=1$ supersymmetric QCD (SQCD), as well as
the quark mass anomalous dimensions in Abelian and non-Abelian
gauge theories.
We show how the accuracy of our previous
$\beta$-function predictions at the four-loop
level may be further improved by using estimators
weighted over negative numbers of flavours (WAPAP's).
The accuracy of the improved four-loop results encourages confidence in
the new five-loop $\beta$-function predictions that we present. However,
the WAPAP
approach does not provide improved results for the anomalous mass
dimension, or for Abelian theories.
\vskip0.6cm
\Date{October 1997}}
\vfill\eject
} 
\pageno=1
\newsec{Introduction}
One of the greatest challenges in QCD is the calculation of
higher orders in perturbation theory. Phenomenologically, these
are important because the relatively large value of $\alpha_s$
at accessible energies implies that many orders of perturbation
theory are required in order to make precise quantitative tests.
Theoretically, one expects the coefficients of the
perturbative series for many QCD quantities to diverge
factorially, and the rates of these divergences may cast
light on issues in nonperturbative QCD, such as the
existence and magnitudes of condensates and higher-twist
effects~\ref\pert{
For recent reviews and references, see:\hfill\break
M. Beneke,
talk at the {\it Fifth
International Workshop on Deep Inelastic Scattering and QCD},
Chicago 1997, hep-ph/9706457;\hfill\break
R. Akhoury and V.I. Zakharov, talks at the {\it QCD~96} and {\it QCD~97}
conferences, Montpellier 1996 and 1997, hep-ph/9610492 and hep-ph/9710257;
\hfill\break
Yu.L.~Dokshitser and B.R.~Webber,
Phys. Lett. {\bf B}404 (1997) 321.}

On the other hand, whilst progress in the exact calculations
of higher-order terms in perturbative QCD series has been startling,
with many new multi-loop results having recently become
available~\ref\calx{See, for example: \hfill\break
T. van Ritbergen, J.A.M. Vermaseren, S.A. Larin and P. Nogueira,
Int. J. Mod. Phys. {\bf C}6 (1995) 513;\hfill\break
A.L. Kataev,
talk at
{\it Second Workshop on Continuous Advances in QCD}, Minneapolis, 1996,
hep-ph/9607426; \hfill\break
see also refs.~[9], [13] and [14].},
existing perturbative techniques may not enable much further
progress in exact calculations to be made in the near future.
Thus various approximate techniques and numerical estimates
may have a useful r\^ole to play. Among these, one may mention
exact calculations of certain
perturbative coefficients in the large-$N_F$ limit, and the
emerging lore of renormalons~\pert. Also of potential use in QCD are
Pad\'e Approximants (PA's), as described in section 2
of this paper, which have previously demonstrated
their utility in applications to problems in condensed-matter physics
and statistical mechanics~\ref\PAcond{M.A. Samuel, G. Li and E. Steinfelds,
Phys. Rev. {\bf D}48 (1993) 869 and
Phys. Lett. {\bf B}323 (1994) 188; \hfill\break
M.A. Samuel and G. Li, Int. J. Th. Phys. 33 (1994) 1461 and Phys. Lett.
{\bf B}331 (1994) 114.}.
In recent years, these have been applied
to obtain successful numerical predictions in various quantum
field theories, including QCD, and justifications for some of these
successes have been found in some mathematical
theorems~\ref\theorems{M.A. Samuel, J. Ellis and M. Karliner, Phys. Rev.
Lett. 74 (1995) 4380; \hfill\break
J. Ellis, E. Gardi, M. Karliner and M.A. Samuel, Phys. Lett. {\bf B}366
(1996) 268 and
Phys. Rev. {\bf D}54 (1996) 6986; \hfill\break
E. Gardi, Phys. Rev. {\bf D}56 (1997) 68; \hfill\break
S.J. Brodsky, J. Ellis, E. Gardi, M. Karliner and M.A. Samuel,
hep-ph/9706467, accepted for publication in Phys. Rev. {\bf D}.} on the
convergence and renormalization-scale invariance of PA's. These
theorems apply, in particular, to perturbative QCD series
dominated by renormalon singularities, and in the large-$\beta_0$ limit.

Based on these theorems,
a new method was introduced \ref\EKS{J.~Ellis, M.~Karliner and
M.A.~Samuel, \plb400 (1997) 176.} for estimating the next-order
coefficients in perturbative quantum field theory series on
the basis of the known
lower-order results and plausible conjectures on the likely
high-order behaviour of the series, as also reviewed in section 2.
This method ``corrects'' the conventional Pad\'e Approximant Prediction (PAP)
of the next term in the series by using an asymptotic error formula,
providing improved predictions that we
call Asymptotic Pad\'e Approximant Predictions (APAP's).

APAP's have already provided successful predictions for
the perturbative coefficients in the subsequent
calculation of the four-loop $\beta$ function in QCD,
as discussed in section 3,
and have also provided interesting results in $N = 1$ supersymmetric QCD
(SQCD)~\ref\jjs{I.~Jack, D.R.T.~Jones and
M.A.~Samuel, Phys. Lett. {\bf B}407 (1997) 143.}.
The purpose of this paper is to provide a more complete
account of these predictions, to show how their accuracy may
be improved in certain cases by a judicious weighting over
negative numbers of flavours $N_F$, and to extend
these predictions
to five loops in QCD in sections 5 and to SQCD in section 6.
We also discuss analogous
predictions for the QCD anomalous quark mass dimension in section 7,
where ``regular'' APAP gives very good results, but
the new weighting method does not improve matters.
In section 8 we consider
Abelian gauge theories,  with less successful results.

Before deriving these predictions, there is a technical issue
which should be clarified, that may also illuminate an
interesting physics point. As a general rule,
$\beta$ functions are scheme-dependent beyond one loop, and
a theory with a single perturbative coupling constant $g$, such as QCD,
is scheme-dependent beyond two loops,
if one considers analytic redefinitions of $g$. In particular,
the QCD $\beta$ function can be transformed to zero beyond two loops,
by making a suitable choice of renormalization scheme~\foot{In fact, it
can even be transformed to
zero beyond {\it one\/} loop by a non-analytic redefinition
of $g$ involving $\ln g$: such redefinitions are associated with the
Wilsonian action in \sic\ theories.}. In our analysis of
QCD, we use the $\overline{MS}$ scheme, and in $N = 1$ SQCD
we favour
the DRED scheme~\foot{We recall that DRED corresponds to minimal
subtraction
in conjunction with regularisation by dimensional reduction.}.
The successes of the APAP procedure
indicate that asymptotia and the convergence of
PAP's are remarkably precocious in these schemes.
In the SQCD case, there exists an alternative scheme
(NSVZ)~\ref\nov{V.~Novikov et al, \npb 229 (1983) 381\semi
V.~Novikov et al, \plb166 (1986) 329\semi
M.~Shifman and A.~Vainstein, \npb 277 (1986) 456\semi
A.~Vainstein, V.~Zakharov and M.~Shifman,
\sjnp 43 (1986) 1028\semi
M.~Shifman, A.~Vainstein and V.~Zakharov \plb 166 (1986) 334.},
associated with the Wilsonian action, in which there
is an all-orders relation
between $\beta_g$ and the quark anomalous dimension $\ga_q$.
The NSVZ scheme
differs perturbatively from DRED
\ref\jjn{I.~Jack, D.R.T.~Jones and C.G.~North, \npb 486 (1997) 479.},
and therefore provides a distinct
test for the APAP method. We compare predictions for $\beta_g$ in
both DRED and NSVZ, finding that they are less compelling in the
latter case: perhaps minimal subtraction schemes are more
amenable to Pad\'e techniques? If so, it would be interesting to
fathom the reason. As already noted, these techniques are not
so successful for the quark mass anomalous  dimension, or for Abelian
theories. Perhaps these instances also provide clues when and why the
Pad\'e magic works.

\newsec{Formalism}
We start by recalling relevant aspects of the formalism
for PA's and APAP's, and establishing our notation.
For a generic perturbative series
\eqn\inta{S(x)=\sum_{n=0}^{N_{max}}
S_nx^n,}
the Pad\'e approximant $[N/M](x)$ is given by~\PAcond\
\eqn\intb{[N/M]= \frakk{a_0 + a_1 x+ \cdots a_N x^N}
{b_0 + b_1 x+ \cdots b_M x^M}}
with $b_0 = 1$, and the other coefficients chosen so that
\eqn\intc{[N/M] = S + O(x^{N+M+1}).}
The coefficient of the $x^{N+M+1}$ term in \intc\ is
the PAP estimate $S_{N+M+1}^{\rm PAP}$ of $S_{N+M+1}$.
If the perturbative coefficients $S_n$ diverge as $n!$
for large $n$, it is possible to show~\theorems\ that
the relative error
\eqn\intd{\delta_{N+M+1}\equiv
{S^{\rm PAP}_{N+M+1} - S_{N+M+1}
\over S_{N+M+1}}}
has the asymptotic form
\eqn\inte{\delta_{N+M+1}\simeq\,{-} {M! {\cal A}^M \over L_{[N/M]}^M} }
as $N\rightarrow\infty$, for fixed $M$, where
\eqn\intf{L_{[N/M]} =N+M+aM+b} and ${\cal A},a,b$ are constants.
This theorem not only guarantees the convergence of the PAP's, but also
specifies the asymptotic form of the corrections.

The idea of APAP's is to fit the magnitude of this asymptotic
correction using the known low-order perturbative coefficients, and apply
the resulting numerical correction to the \naive\ PAP's.
In the applications discussed in this paper,
we work with $[0/1], [1/1]$ and $[2/1]$ PA's, so that $M=1$ throughout.
For example, four-loop predictions are obtained as follows.
In the case $N_{max}=2$, the $[1/1]$ Pad\'e leads to the  \naive\ PAP
$S_3^{\rm PAP} = S_2^2/S_1$. The improved APAP estimate
is then given by
\eqn\intg{
S_3^{\rm APAP} = \frakk{S_3^{\rm PAP}}{1 + \delta_3,}}
where, motivated by its appropriateness in $\phi^4$ field theory, we
choose $a+b=0$ in the QCD application discussed in the next
section, and ${\cal A}$ is then determined by comparing $S_2$ to
$S_2^{\rm PAP} = S_1^2/S_0$. Alternatively, we could have chosen a  value
of
${\cal A}$ and determined $a+b$ from $\delta_2$. However, as we shall see,
when we go to
five loops, knowledge of $\delta_2$ and $\delta_3$ enables
us to fit both ${\cal A}$ and $a+b$ simultaneously.

\newsec{Application to the Four-Loop $\beta$ Function in QCD}

The APAP method was applied in~\EKS\ to estimate the four-loop
QCD $\beta$ function coefficient $\beta_3$, on the basis of
the lower-order terms
\eqn\qcdbII{\eqalign{
\beta_0&=\frak{11}{3}C_A-\frak{4}{3}T_FN_F,\cr
\beta_1&=\frak{34}{3}C_A^2-4C_FT_FN_F-\frak{20}{3}C_AT_FN_F,\cr
\beta_2&=\frak{2857}{54}C_A^3+N_F\Bigl[2C_F^2T_F-\frak{205}{9}C_FC_AT_F
-\frak{1415}{27}C_A^2T_F\Bigr]\cr&
+N_F^2\Bigl[\frak{44}{9}C_FT_F^2+\frak{158}{27}
C_AT_F^2\Bigr],}}
known before the appearance of
the explicit four-loop calculation~\ref\larin{T.~van Ritbergen,
J.A.M.~Vermaseren
and S.A.~Larin, \plb400 (1997) 379.}.
The quadratic Casimir coefficients $C_A$ and $C_F$ for the adjoint and
fundamental representations are given for the case of $SU(N_C)$ by
\eqn\qcdc{
C_A=N_C,\qquad C_F={N_C^2-1\over{2N_C}},}
and we assume the standard normalisation so that $T_F=\frak{1}{2}$.
We denote by $N_A$ the
number of group generators, so that for $SU(N_C)$ we have $N_A=N_C^2-1$.

We recall that $\beta_3$ is a
polynomial in the number of flavours $N_F$:
\eqn\inth{
\beta_3 = A_3 + B_3 N_F + C_3 N_F^2 + D_3 N_F^3,}
where $D_3 = 1.499$ (for $N_C=3$)
was already known from large-$N_F$ calculations.
To justify applying the estimate \inte, we assume that the $\beta_n \sim n!$
for large $n$, as discussed in~\EKS.
The predictions for $A_3, B_3, C_3$ resulting from
fitting the APAP results for $0 \le N_F \le 4$ to a
polynomial of
the form \inte\ are compared to the exact results in
the first columns of Table~I.

The exact four-loop coefficient of the QCD $\beta$ function for $N_C$ colours
is taken from the calculation of~\larin, which was published
after the APAP estimate:
\eqn\betaIII{\eqalign{
\beta_3&=C_A^4\left(\frak{150653}{486}-\frak{44}{9}\zeta_3\right)
+{d_A^{abcd}d_A^{abcd}\over{N_A}}\left(-\frak{80}{9}
+\frak{704}{3}\zeta_3\right)\cr
&+N_F\Bigl[C_A^3T_F
\left(-\frak{39143}{81}+\frak{136}{3}\zeta_3\right)+C_A^2C_FT_F\left
(\frak{7073}{243}-\frak{656}{9}\zeta_3\right)\cr&
+C_AC_F^2T_F
\left(-\frak{4204}{27}+\frak{352}{9}\zeta_3\right)+46C_F^3T_F
+{d_F^{abcd}d_A^{abcd}\over{N_A}}\left(\frak{512}{9}-\frak{1664}{3}
\zeta_3\right)\Bigr]\cr&
+N_F^2\Bigl[C_A^2T_F^2\left(\frak{7930}{81}+\frak{224}{9}\zeta_3\right)
+C_F^2T_F^2\left(
\frak{1352}{27}-\frak{704}{9}\zeta_3\right)
+C_AC_FT_F^2\left(\frak{17152}{243}+\frak{448}{9}\zeta_3\right)\cr&
+{d_F^{abcd}d_F^{abcd}\over{N_A}}\left(-\frak{704}{9}+\frak{512}{3}
\zeta_3\right)\Bigr]
+N_F^3\Bigl[\frak{424}{243}C_AT_F^3+\frak{1232}{243}C_FT_F^3\Bigr],
\cr}}
where $\zeta_3\equiv \zeta(3)= 1.2020569\cdots$. The
quartic Casimir coefficients in \betaIII\ are given for $SU(N_C)$ by
\eqn\qcdd{\eqalign{
d_A^{abcd}d_A^{abcd}=&{N_C^2(N_C^2-1)(N_C^2+36)\over{24}},
\qquad d_F^{abcd}d_A^{abcd}={N_C(N_C^2-1)(N_C^2+6)\over{48}},\cr
d_F^{abcd}d_F^{abcd}&={(N_C^2-1)(N_C^4-6N_C^2+18)\over{96N_C^2}}.
\cr}}
For $N_C=3$ one obtains
\eqn\qcdf{\eqalign{
\beta_3&\approx29243.0-6946.30N_F+405.089N_F^2+1.49931N_F^3,}}
whereas $\beta_3$ is given by
the coefficients shown in Table I when one omits the
quartic Casimir contributions.

These quartic Casimir terms appear for the
first time at four-loop order. They are analogous to the
light-by-light scattering terms in $(g - 2)_{\mu}$, and PA-based
techniques cannot estimate them on the basis of lower-order terms
with different group-theoretical factors. Such terms are known
to be important in $(g - 2)_{\mu}$, but were relatively unimportant
in previous perturbative QCD applications. In the case of $\beta_3$,
they turn out to be about 15 to 20 \% for small $N_F$, but
are non-negligible for $N_F \sim 5$.
Setting these terms aside, the agreement between
the predictions of~\EKS\ and the exact results of~\larin\ is
remarkable. The predictions we present in the rest
of this paper should all be understood as applying to
perturbative coefficients without the higher-order analogues of
such quartic Casimir terms.

\vskip3em
\vbox{
\begintable
     | \hb{APAP}   |\hb{EXACT}|\hb{\% DIFF} |\mb|\hb{WAPAP}  |\hb{\% DIFF}\cr
A_3  | 23,600(900) | 24,633   | -4.20(3.70) |\mb| 24,606     | -0.11      \cr
B_3  | -6,400(200) | -6,375   | -0.39(3.14) |\mb|-6,374      | -0.02      \cr
C_3  | 350(70)     | 398.5    | -12.2(17.6) |\mb| 402.5      | -1.00      \cr
D_3  | \hb{input}  | 1.499    |    -        |\mb|\hb{input}  | -
\endtable
\bigskip
\in
{\it \noindent Table~I: Exact four-loop results for the QCD
$\beta$ function, compared with the original APAP's in the first column,
and improved APAP's obtained from a weighted average over negative
$N_F$ (WAPAP), as discussed in the text.
The numbers in parenthesis are the error estimates from~\EKS.}
\bigskip \out}

Following \EKS,
the same APAP method was applied in \jjs\ to estimate the four-loop
$\beta$ function in SQCD. The agreement with known results was
again encouraging, and the APAP provided a
prediction $\alpha\approx 2.4$ for the unknown
constant \jjn\ in the four-loop SQCD $\beta$ function,
as discussed also in section~5.

\newsec{Weighted APAP's in QCD}

Before going on to make new predictions for QCD and SQCD at the
five-loop level, we first draw attention to a refinement that
offers an improvement on APAP's in the four-loop QCD case.
As can be seen in Table I, the signs of the coefficients
$A_3, B_3, C_3$ alternate. A corollary of this is that the
APAP predictions for $N_F \sim 5$ are sensitive to
cancellations and relatively inaccurate. Conversely, the
numerical analysis is relatively stable for (fictitious)
$N_F < 0$. We have observed empirically that more
accurate predictions for the coefficients
$A_3, B_3$ and $C_3$ are obtained
if one makes polynomial fits for some range of
{\it negative} values of $N_F$. Is there some systematic
procedure that exploits this observation?
The following is one method we have explored.

We choose a range $-N_F^{max} \le N_F \le 0$ over which
we fit values of $\cal A$ using the APAP formulae of the previous
section, and we determine the arithmetic mean of the
corresponding values of $\cal A$. We use this mean value of
$\cal A$ to estimate $\beta_3$ for each of the chosen values of
$N_F$, and fit to the polynomial form \inth . We hypothesize
that the most accurate results for the
coefficients $A_3, B_3, C_3$ may be obtained when they
contribute with equal weights to the fit: certainly, one
cannot expect that any coefficient that has a small weight in the fit
will be estimated reliably. For given $N_F^{max}$, the overall
weights in the fit are $A_3, \ B_3 N_F^{max}/2$ \ and \
$C_3 N_F^{max} (2N_F^{max} + 1) /6$. We then estimate $B_3$ as follows.
We take the two values of $B_3$ corresponding to the
values of $N_F^{max}$ for which the $A_3$ and $B_3$
weights are most nearly equal. Let us call these values of $B_3$,
$B_3^{(1)}$ and $B_3^{(2)}$, and the corresponding weights
$B_3^{W(1)}$ and $B_3^{W(2)}$. Our prediction for $B_3$ is then
\eqn\bweight{
B_3 = \frakk{\Delta_2 B_3^{(1)} + \Delta_1 B_3^{(2)}}{\Delta_1 + \Delta_2},
}
where $\Delta_{1,2}= |B_3^{W(1,2)}-A_3^{W(1,2)}|$. We estimate $C_3$
in similar fashion. Both the $B_3$ calculation and the $C_3$ calculation
yield a result for $A_3$, obtained as in \bweight\ : we
take as our prediction for $A_3$ the mean of these two  values.

Table I shows in the column labelled WAPAP the results we obtain using
this procedure. We see that the latter are significantly
more accurate than the ones obtained using the APAP's for $0 \le N_F \le
4$. The values of $N_F^{max}$ selected by WAPAP are
$7,8$ for $B_3$ and $13,14$ for $C_3$.

Table II compares the WAPAP predictions obtained in this way
with the known exact results (omitting quartic Casimir contributions)
in QCD for various values of $N_C$.
The agreement is certainly impressive, even compared with the
APAP results shown in Table I. Since the numerical value of the
coefficient $C_3$ is relatively small, corresponding (in the case $N_C=3$)
to the relatively large value $N_F^{max} = 14$ mentioned above, it
is perhaps not surprising that the percentage error in the estimate
of this coefficient is larger than for either $A_3$ or $B_3$.

Figure 1 displays graphically our resulting predictions for $\beta_3$,
as a function of $N_F$ for the most interesting case $N_C = 3$.
We plot the percentage relative errors obtained using various APAP-based
estimation schemes: naive APAP's fitted with positive $N_F \le 4$
(diamonds),
naive APAP's fitted with negative $N_F \ge -4$, WAPAP's compared to the
exact value of $\beta_3$ including quartic Casimir terms, and
WAPAP's compared to $\beta_3$ without quartic Casimir terms (crosses). We
see that the latter are the most accurate for $\beta_3$ in QCD.
In Figure 2 we show the error in the  WAPAP prediction for $\beta_3$
as a function of $N_F$, for $N_C=$3, 4, 5, 6, 7 and 10,
once again omitting quartic Casimir terms from the exact result.
The accuracy of these predictions is our best evidence for
believing in the utility of the WAPAP method.

To anticipate the obvious question: we have explored whether
this WAPAP procedure gives significantly better results than
the conventional APAP's for the other perturbative series
considered in this paper, namely the SQCD $\beta$ function
and the anomalous dimension of the quark mass. As we discuss in sections
7 and 8, the remarkable success of the method at four loops
is not repeated for other cases, but
there is distinct evidence (provided by large-$N_F$-expansion results)
that
WAPAP leads to more reliable predictions at five loops.
However, we feel that the results in Tables I and II already
provide ample motivation for the QCD WAPAP calculation of $\beta_4$
described in the next section.

\bigskip
\vbox{ 
\noncenteredtables
\line{\kern14.2em
\begintable
\kern0.27em\hb{WAPAP} \kern0.77em|
\kern1.04em\hb{exact}\kern1.04em|
\kern-0.2em\hb{\% error}\kern-0.56em
\endtable
}
\centeredtables
\tablewidth=-\maxdimen

\thicksize=0.0pt
\begintable
\bb &N_C=2 & \multispan{3} \mystrut \hfill      & \bb \cr
\bb | A_3    | 4.88 \times 10^3  | 4,866     |  0.42       | \bb \cr
\bb | B_3    | -1.86 \times 10^3  | -1,854     |  0.48       | \bb \cr
\bb | C_3    |  174  | 170.5      |  2.0       | \bb \cr
\bb &N_C=3 & \multispan{3} \mystrut \hfill      & \bb \cr
\bb | A_3    | 2.467 \times 10^4  | 24,633     |  0.13       | \bb \cr
\bb | B_3    | -6.383 \times 10^3  | -6,375     |  0.13       | \bb \cr
\bb | C_3    |  405  | 398.5      |  1.6       | \bb \cr
\bb &N_C=4 & \multispan{3} \mystrut \hfill      & \bb \cr
\bb | A_3    | 7.790 \times 10^4  | 77,852     |  0.06       | \bb \cr
\bb | B_3    | -1.521 \times 10^4 | -15,210     |  0.03      | \bb \cr
\bb | C_3    |  729  | 717.2      |  1.6       | \bb \cr
\bb &N_C=5 & \multispan{3} \mystrut \hfill      & \bb \cr
\bb | A_3    | 1.901 \times 10^5 | 190,068    |  0.04       | \bb \cr
\bb | B_3    | -2.976 \times 10^4 | -29,800    | -0.12      | \bb \cr
\bb | C_3    |  1.14 \times 10^3 | 1,127      |  1.6       | \bb \cr
\bb &N_C=6 & \multispan{3} \mystrut \hfill      & \bb \cr
\bb | A_3    | 3.943 \times 10^5 | 394,125    |  0.03       | \bb \cr
\bb | B_3    | -5.149 \times 10^4 | -51,580     | -0.17       | \bb \cr
\bb | C_3    | 1.65 \times 10^3  | 1,627.5    |  1.6       | \bb \cr
\bb &N_C=10 & \multispan{3} \mystrut \hfill      & \bb \cr
\bb | A_3    |\,\,3.043 \times 10^6| 3,041,089  | 0.05       | \bb \cr
\bb | B_3    | -2.388 \times 10^5 | -239,384     | -0.25       | \bb \cr
\bb | C_3    |  4.62 \times 10^3 | 4,540      |  1.7      | \bb \cr
\bb & \multispan{4}                             & \bb
\endtable
\thicksize=0.7pt
\bigskip
\in
{\it \noindent Table~II:
Comparison of WAPAP and exact results for the exact 4-loop $\beta$
function in QCD (omitting quartic Casimir terms),
for various values of $N_C$.}
\bigskip
\out
}
\vfill\eject
\line{\hfill}
\vskip5em
\FIG{Predictions for $\beta_3$, as function of $N_F$,
for $N_C=3$. The percentage relative errors are obtained
using various APAP-based estimation schemes:
naive APAP's fitted with positive $N_F \le 4$
(diamonds),
naive APAP's fitted with negative $N_F \ge -4$, WAPAP's compared to the
exact value of $\beta_3$ including quartic Casimir terms, and
WAPAP's compared to $\beta_3$ without quartic Casimir
terms (crosses).}{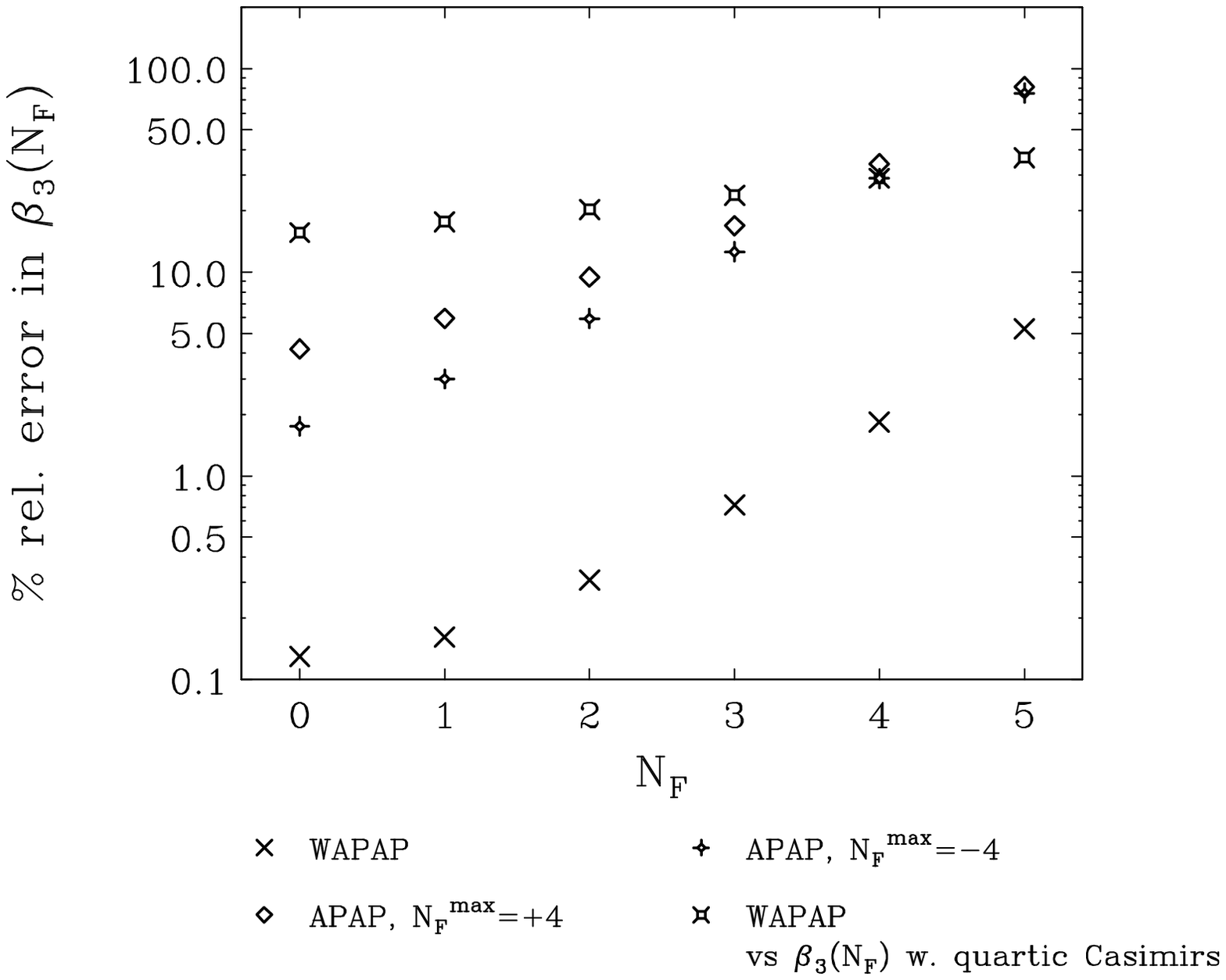}{14truecm}{1.5truecm}
\figlabel\FigI
\line{\hfill}
\bigskip
\vfill\eject
\medskip
\FIG{The percentage relative errors in the WAPAP prediction
for $\beta_3$ (compared to the exact result with quartic Casimir terms
omitted), plotted vs. $N_F$
for $N_C=$3, 4, 5, 6, 7, 10.}{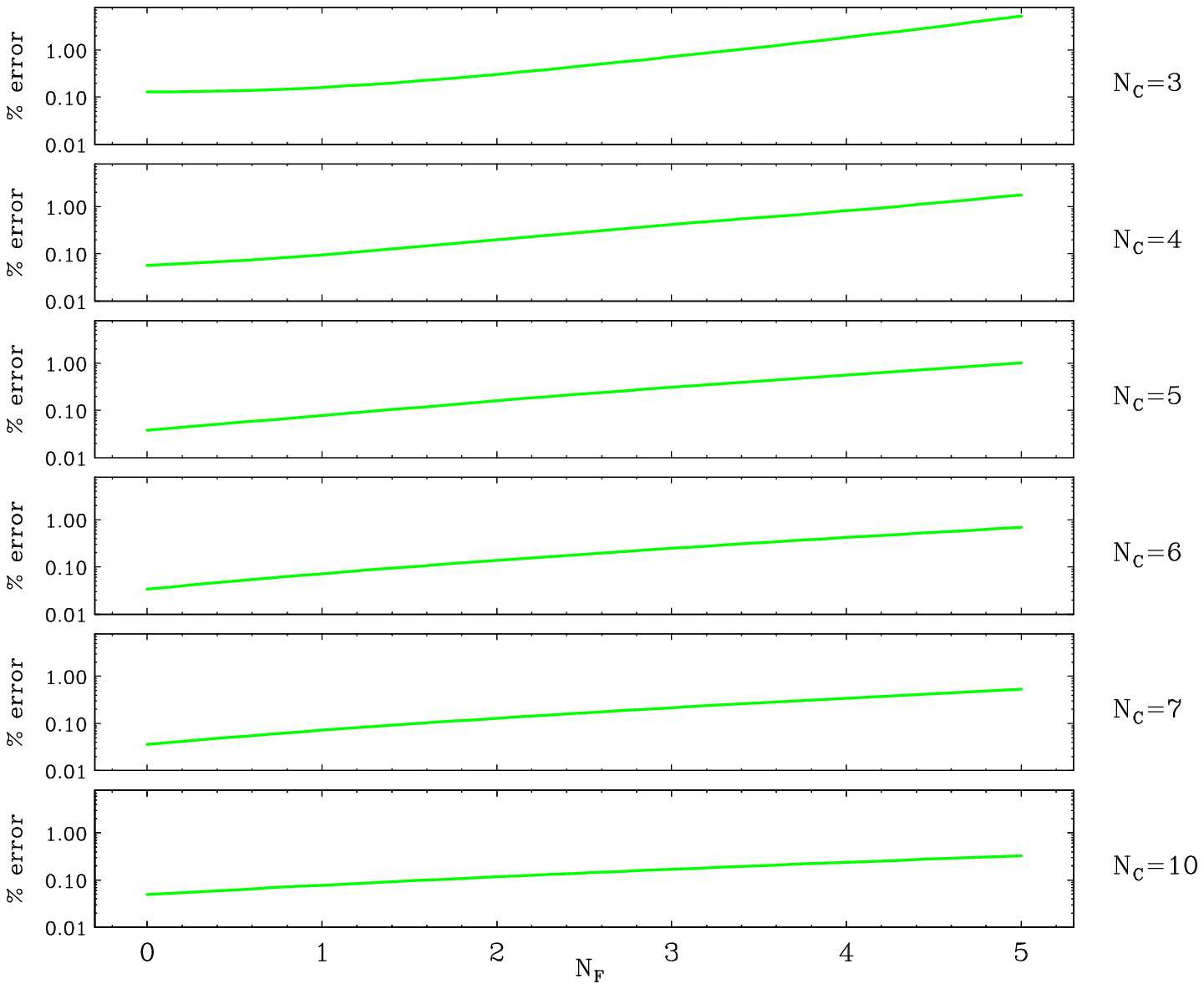}{22truecm}{-12truecm}
\figlabel\FigIa
\medskip

\newsec{Five-Loop Predictions in QCD}

We now outline the application of the APAP method to estimate the five-loop
$\beta$ function coefficients
$\beta_4$ in QCD, using our knowledge of the corresponding
$\beta_0$ to $\beta_3$. The standard $[2,1]$ Pad\'e leads to the estimate
\eqn\inti{
\beta_4^{\rm PAP}={\beta_3^2\over{\beta_2}}.}
This is then corrected in a similar fashion to Eq.~\intg:
\eqn\intj{\beta_4^{\rm APAP}={{\beta_4^{\rm PAP}}\over{1+\delta_4}},}
where, according to Eqs.~\inte, \intf,  $\delta_4$ is given
asymptotically by \eqn\intk{
\delta_4=-{{\cal A}\over{L_{[2/1]}}}=-{{\cal A}\over{3+a+b}}.}
To estimate $\delta_4$ we therefore need to know both $\cal A$ and $a+b$.
These can be
deduced from the lower-order relative errors $\delta_2$ and $\delta_3$, as
defined in \intd\ for which we use the asymptotic estimates \inte:
\eqn\intl{
{{\cal A}\over{\delta_2}}=-(1+a+b),\qquad {{\cal
A}\over{\delta_3}}=-(2+a+b) } from which we obtain $\cal A$ and
$a+b$~\foot{The fitted value of $a+b$ is not necessarily close
to the value zero assumed in the estimate of $\beta_3$ in QCD.}.

We now calculate the WAPAP for the five-loop QCD $\beta$ function,
which we parametrise as
\eqn\qcdj{
\beta_4=A_4+B_4N_F+C_4N_F^2+D_4N_F^3+E_4N_F^4.}
Once again we can input the coefficient of the highest power in $N_F$,
which is given in this case by
\ref\Gracey{J.A. Gracey, \plb373 (1996) 178.}:
\eqn\qcdja{
E_4= -4T_F^4\left[(288\zeta(3)+214)C_F+(480\zeta(3)-229)C_A\right]/243\ .}
using which we obtain the five-loop results shown in Table~III.

\vskip 3em
\vbox{
\begintable
N_C       |  2        |     3       |       4      |    5
|  10            \cr
A_4 \wQ       |  1.48\times10^5  |   7.59\times10^5  |  2.77\times10^6   |
  7.92\times10^6|  2.31\times10^8\cr
A_4 \woQ  |  6.41\times10^4  |   4.88\times10^5   |  2.06\times10^6  |
6.28\times10^6|  2.01\times10^8 \cr
B_4 \wQ | -5.51\times10^4   |    -2.19\times10^5 | -6.39\times10^5
|  -1.50\times10^6| -2.28\times10^7   \cr
B_4 \woQ   | -3.04\times10^4  |    -1.56\times10^5 | -4.97\times10^5
|  -1.22\times10^6 | -1.95\times10^7   \cr
C_4 \wQ      | 6.96\times10^3   |  2.05\times10^4     | 4.68\times10^4
|   9.00\times10^4| 7.07\times10^5   \cr
C_4 \woQ     | 4.69\times10^3   |  1.64\times10^4     | 3.93\times10^4
|   7.72\times10^4| 6.23\times10^5   \cr
D_4 \wQ  | -21.8    | -49.8      | -89.8    |  -142
| -575      \cr
D_4  \woQ  | -28.3    | -60.5      | -105    |  -163
| -640     \cr
E_4(\hb{input})| -1.15     | -1.84       | -2.51        |   -3.17
|  -6.43
\endtable
\bigskip
\in
{\it \noindent Table~III: WAPAP's for the five-loop QCD $\beta$ function,
calculated both with (w.~Q) and without (w/o~Q) the
four-loop quartic Casimir terms in $\beta_3$. The values of $N_F^{max}$
used range between 5 and  117 in the w. Q case, and between 4 and 108
in the w/o~Q case, being largest for large $N_C$ and for $D_4$.}
\out
}
\bigskip
Notice that in Table~III we include results corresponding to
both the inclusion (w. Q) and the omission (w/o Q) of the quartic Casimir
contributions
to the four-loop coefficients, obtained from \betaIII.
The former (latter) results
should of course be compared with contributions including (excluding)
such terms at five loops when (and if) such results become available.
Of course, at five-loop order we may expect to encounter new higher-order
Casimir terms,
which should in any event be omitted in the comparison. We can only hope
that such contributions are relatively unimportant, which is the case for
the quartic terms in $\beta_3$ for small $N_F$.
We anticipate that the percentage errors of the w/o Q estimates
of the non-quartic terms in the coefficients are likely to be the
smallest, whereas
the best estimate of the full coefficients may be provided by the
w. Q estimates.

We show below the results obtained if we choose not to input the
value of $E_4$, but rather predict that as well. As can be
seen, the results
for $A_4$, $B_4$ and $C_4$, in particular,  are very stable. Moreover,
the prediction for $E_4$ is
encouragingly close to the true value, considering the extreme smallness
of
$E_4$ compared to $A_4$.
\vskip 3em
\vbox{
\begintable
N_C       |  2        |     3       |       4      |    5
|  10            \cr
A_4 \wQ  |  1.45\times10^5  |   7.51\times10^5   |  2.75\times10^6  |
  7.87\times10^6
|  2.30\times10^8\cr
A_4 \woQ   |  6.38\times10^4  | 4.85\times10^5     |  2.05\times10^6  |
6.24\times10^6
|  2.00\times10^8\cr
B_4 \wQ| -5.53\times10^4   |    -2.20\times10^5 | -6.41\times10^5
|  -1.51\times10^6  | -2.29\times10^7   \cr
B_4 \woQ | -3.05\times10^4  |    -1.57\times10^5 | -4.99\times10^5
|  -1.22\times10^6
| -1.96\times10^7\cr
C_4  \wQ | 6.72\times10^3   |  1.97\times10^4    | 4.50\times10^4
|   8.66\times10^4| 6.81\times10^5   \cr
C_4  \woQ | 4.52\times10^3   |  1.58\times10^4    | 3.79\times10^4   |
7.43\times10^4| 5.99\times10^5   \cr
D_4  \wQ | -28.3    | -93.8      | -226    |  -389
| -1,730   \cr
D_4 \woQ    | -72.7    | -163     | -287   |  -446
| -1,750  \cr
E_4 \wQ | -0.974    | -2.03       | -3.07        |   -4.06
|  -8.73\cr
E_4 \woQ| -1.61    | -2.56       | -3.45        |   -4.33
|  -8.64
\endtable
\bigskip
\in
{\it \noindent Table~IV: WAPAP's for the five-loop QCD $\beta$ function,
calculated with and without the four-loop quartic Casimir terms, but
without inputting the known exact values of $E_4$. It is encouraging to
compare the output values with the last row in Table III.
The values of $N_F^{max}$ used range between 5 and 81 in the w.~Q case,
and between 4 and 104 in the w/o~Q case.} \bigskip
\out
}

It is not possible to state precise errors for
the type of prediction discussed in this paper. We gave
in~\EKS\ certain estimates of the uncertainties, which
turned out to be in the right ballpark if quartic
Casimir terms are omitted in the comparison, as
reported in Table~I. The appearance of such new
quartic terms is characteristic of the type of
theoretical `systematic error' that cannot be
foreseen. In the case of our $\beta_4$ predictions
in QCD, we draw the reader's attention to the
differences between the w. Q and w/o Q entries in Table~III,
and to the differences between these and the corresponding
entries in Table~IV, obtained without using the known values
of $E_4$ as inputs. The most accurate estimates
of the full coefficients are likely
to be the w. Q entries in Table~III, but the uncertainties
are unlikely to be smaller than these differences.

\newsec{Five-Loop Predictions in $N = 1$ Supersymmetric QCD}

We begin with the SQCD $\beta$ function
in the DRED regularisation scheme, where
the first four coefficients are given
by~\jjn:
\def\eqstrut{\vrule height 4.0ex depth 2.0ex width 0pt}
\eqna\sqcdb$$\eqalignno{
\beta_0 &= 3N_C - N_F, \eqstrut &\sqcdb a\cr
\beta_1 &= 6N_C^2- \left[4N_C-{2\over {N_C}}\right]N_F \eqstrut &\sqcdb b\cr
\beta_2 &= 21N_C^3 - \left[21N_C^2-{2\over{N_C^2}}-9\right]N_F
-\left[{3\over{N_C}}-4N_C\right]N_F^2 &\sqcdb c\cr
\beta_3 &= A_3 + B_3N_F + C_3N_F^2 +D_3N_F^3 \eqstrut&\sqcdb d\cr}$$
where $N_C$ is the number of colours, and
\eqn\sqcdc{\eqalign{
A_3 &= (6+36\alpha)N_C^4\cr
B_3 &=  -36(1+\alpha)N_C^3+(34+12\alpha)N_C+\frakk{8}{N_C}
+\frakk{4}{N_C^3}\cr
C_3 &= \left(\frakk{62}{3}+2\kappa+8\alpha\right)N_C^2-\frakk{100}{3}
-4\alpha-\frakk{6\kappa-20}{3N_C^2}\cr
D_3 &= \frakk{2}{3\,N_C}.\cr}}
Here $\kappa=6\zeta_3$ and
$\alpha$ is a constant which has not yet been calculated exactly.
Notice that there are no quartic Casimir contributions in the
SQCD case~\foot{Their absence may be understood as a consequence of the fact
that the $\beta$ function vanishes beyond one loop for an arbitrary $N=2$
supersymmetric theory. We are unable, however,
 to comment on the possible appearance of quartic and higher-order
Casimir terms at the five-loop level.}. The APAP
method was used in an earlier paper \jjs\
to obtain the estimate $\alpha\approx2.4$.

Proceeding now to five loops, we write
\eqn\sqcdd{
\beta_4=A_4+B_4N_F+C_4N_F^2+D_4N_F^3+E_4N_F^4.}
As in the QCD case we can input the
true value of $E_4$ provided by a recent large $N_F$
calculation~\ref\jjfn{P.M.~Ferreira, I.~Jack, D.R.T.~Jones and C.G.~North
hep-ph/9705328.}, and given by
\eqn\sqcdea{E_4 = -\left[2N_C\zeta_3 - (1+2\zeta_3)/(2N_C)\right].}
We choose to calculate the WAPAP predictions
both with and without
this input. This also enables us to explore the sensitivity of the
resulting
prediction for $E_4$ to variations in $\alpha$.
Assuming $\alpha=2.4$, we obtain the results shown in Table V, whereas
the results with the known values of $E_4$ not input are shown in
Table~VI. The qualitative agreement between the
predicted values of $E_4$
in the last row of Table VI and the exact values in Table V is
good.  We note that the WAPAP process is crucial
for this agreement, in that
the output $E_4$ is quite sensitive to the value of $N_F^{max}$ used,
which is fixed by the WAPAP
criterion. We see that the output values of $A_4$, $B_4$, $C_4$ and $D_4$
are quite
stable, which is perhaps to be expected in view of the small
numerical values of $E_4$.
The differences between the results obtained with and without the
input exact value of $E_4$ provide some indication of the uncertainty in
the
predictions. We expect, naturally, the case with input $E_4$ to be the more
accurate.

\vbox{ 
\vskip3em
\vbox{
\begintable
N_C       |  2         |     3      |       4   |   5       | 10         \cr
A_4       | 1.48\times10^4      | 1.13\times10^5      | 4.78\times10^5
| 1.46\times10^6    | 4.69\times10^7\cr
B_4       |-1.05\times10^4    | -5.85\times10^4       | -1.91\times10^5
| -4.72\times10^5| -7.70\times10^6  \cr
C_4       | 3.25\times10^3     | 1.29\times10^4      | 3.21\times10^4
| 6.42\times10^4     |5.29\times10^5        \cr
D_4       | -109 | -307   | -583   | -936  | -3.87\times10^3      \cr
E_4(\hb{input)}| -3.96 | -6.64 | -9.19| -11.7 | -23.9
\endtable
\bigskip
\leftskip = 25 pt\rightskip = 10pt
{\it \noindent
Table~V: WAPAP's for the five-loop SQCD $\beta$ function,
assuming $\alpha=2.4$. The values of $N_F^{max}$ used range
between 3 and 37.}
\out
}
\vskip3em
\vbox{
\begintable
N_C       |  2         |     3      |       4   |   5       | 10         \cr
A_4       | 1.46\times10^4      | 1.12\times10^5   |  4.73\times10^5
| 1.45\times10^6 |4.64\times10^7\cr
B_4       |-1.04\times10^4   | -5.87\times10^4       | -1.91\times10^5
| -4.74\times10^5|-7.73\times10^6\cr
C_4       | 3.16\times10^3   | 1.25\times10^4      | 3.11\times10^4
| 6.21\times10^4  |5.12\times10^5       \cr
D_4       | -134 | -400   | -767   | -1.24\times10^3  | -5.12\times10^3      \cr
E_4| -2.44| -4.53 | -6.33| -8.03 | -16.1
\endtable
\bigskip
\leftskip = 25 pt\rightskip = 10pt
{\it \noindent
Table~VI: WAPAP's for the five-loop SQCD $\beta$-function,
again assuming $\alpha=2.4$, but without the exact values of
$E_4$ as input. The values of $N_F^{max}$ used range between
4 and 61.}
\bigskip\out
}
} 

%
\medskip
\FIG{The WAPAP result for $E_4$ plotted against $\alpha$,
for $-3 < \alpha < 3$.}{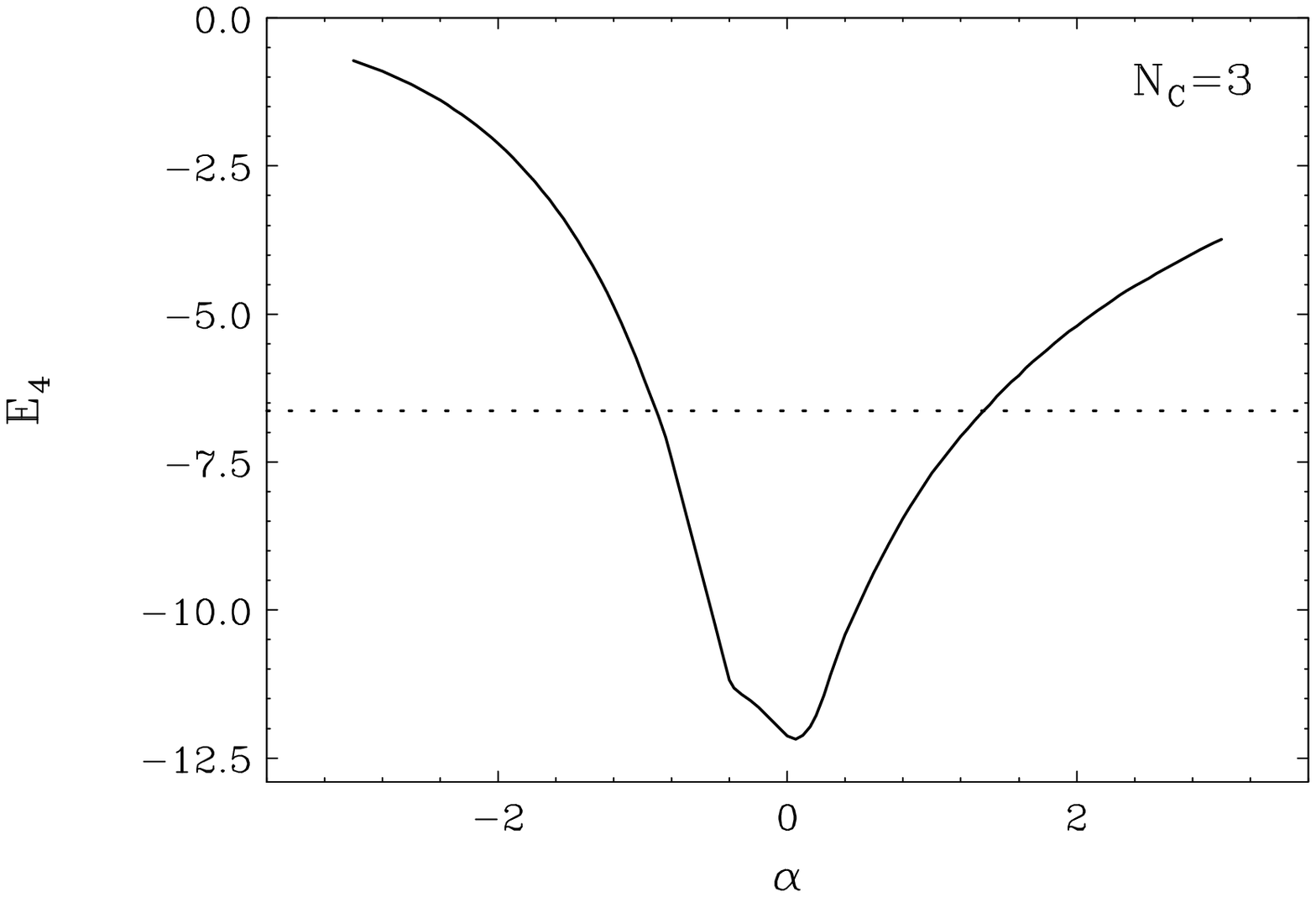}{14truecm}{0truecm}
\figlabel\FigII
\medskip

The value $\alpha=2.4$ used above was itself based on an APAP
calculation~\jjs.  It behoves us, therefore, to explore the sensitivity of
our results  to the precise value of $\alpha$. In Fig.~3 we plot the
WAPAP result for $E_4$ against $\alpha$, for $-3 < \alpha < 3$. We
see that for this range there are two values of $\alpha$ corresponding
to $E_4 = E_4^{\rm exact}$, namely $\alpha\approx -0.9$ and
$\alpha\approx 1.4$.  Given the fact that in general we would expect $E_4$
to be the least-well determined coefficient, we consider this result to be
reasonably consistent with our previous
prediction that $\alpha\approx 2.4$. It should be noted that our
predictions for $A_4\cdots D_4$ are also sensitive to
the precise value of $\alpha$.

We turn now to the alternative NSVZ prescription for the SQCD
$\beta$-function, given by the following
exact formula~\nov\ which relates $\beta_g$ to the
quark anomalous dimension, $\gamma_q$:
\eqn\russa{\beta_g^{\NSVZ} =
-{{g^3}\over{\lf}}\left[ {{N_F-3N_C - 2N_F\gamma_q^{\NSVZ}}
\over{1- 2N_Cg^2{(\lf)}^{-1}}}\right],}
Note the overall minus sign, in accordance with our conventions here.
Using \russa\ and the result for $\gamma_q^{\NSVZ}$
given in \jjn, we obtain:
\def\eqstrut{\vrule height 4.0ex depth 2.0ex width 0pt}
\eqna\svz$$\eqalignno{
\beta_0 &= 3N_C - N_F, \eqstrut &\svz a\cr
\beta_1 &= 6N_C^2- \left[4N_C-{2\over {N_C}}\right]N_F \eqstrut &\svz b\cr
\beta_2 &= 12N_C^3 - \left[12N_C^2-{2\over{N_C^2}}-6\right]N_F
-\left[{2\over{N_C}}-2N_C\right]N_F^2 &\svz c\cr
\beta_3 &= A_3 + B_3N_F + C_3N_F^2 +D_3N_F^3 \eqstrut&\svz d\cr}$$
where
\eqn\svzb{\eqalign{
A_3 &= 24N_C^4\cr
B_3 &=  -40N_C^3+30N_C-\frakk{2}{N_C}
+\frakk{4}{N_C^3}\cr
C_3 &= \left(2\kappa+14\right)N_C^2-24-\frakk{2\kappa-10}{N_C^2}\cr
D_3 &= 2N_C-\frakk{2}{N_C}.\cr}}
In this case there is no undetermined parameter $\alpha$: we know~\jjn\
$\ga_q^{\NSVZ}$ through three loops and hence $\beta_g^{\NSVZ}$
through four loops.

It is possible to argue~\ref\jj{I.~Jack and D.R.T.~Jones, hep-ph/9707278.}
on the basis
of the nature of the coupling-constant redefinition connecting the two
schemes  that $\ga_q^{\DRED}$ and $\ga_q^{\NSVZ}$
are the same at leading order in $N_F$. Hence, if
as before we write
\eqn\svzc{
\beta_4=A_4+B_4N_F+C_4N_F^2+D_4N_F^3+E_4N_F^4,}
we can find $E_4$, as we did
in the DRED case, from the large-$N_F$ results in \jjfn. The result is:
\eqn\svzd{
E_4=2\left[1-2\zeta (3)\right](N_C-1/N_C).}
We also have, as is evident from \russa, that $A_4 = 48N_C^5$,
providing an additional check on our calculation~\foot{We could, of
course, input {\it both\/} $A_4$
and $E_4$, but we choose instead to compare
the WAPAP results for all the five--loop
coefficients with the corresponding ones with
$E_4$ input.}. Our WAPAP results are shown in the Tables~VII and VIII,
for the cases with and without $E_4$ input. Also shown in the second
row of Table~VIII are the exact results for $A_4$.

\vskip3em
\vbox{
\begintable
N_C       |  2         |     3      |       4   |   5       | 10         \cr
A_4       | 1.68\times10^3    | 1.04\times10^4   |  4.44\times10^4
| 4.99\times10^5|4.42\times10^6\cr
B_4       |-1.25\times10^3    | -7.87\times10^3       | -2.63\times10^4
| -6.56\times10^4|-1.08\times10^6\cr
C_4       | 750    | 3.11\times10^3      | 7.87\times10^3    | 1.58\times10^4
|1.32\times10^5       \cr
D_4       | -6.0 |-90.1    | -163   | -516  | -938      \cr
E_4 (\hb{input})| -4.21| -7.49 | -10.5 | -13.5 | -27.8
\endtable
\bigskip
\leftskip = 25 pt\rightskip = 10pt
{\it \noindent
Table~VII: WAPAP's for the five-loop NSVZ $\beta$ function,
with the exact values of $E_4$ used as input. The values of $N_F^{max}$
used range between 3 and 26.}
\out
}

\vskip3em
\vbox{
\begintable
N_C       |  2         |     3      |       4   |   5       | 10         \cr
A_4       | 1.49\times10^3      | 1.05\times10^4   |  4.33\times10^4
| 1.42\times10^5 |4.45\times10^6\cr
A_4(\hb{exact}) | 1.536\times10^3  | 1.166\times10^4 |  4.915\times10^4
 | 1.500\times10^5 |4.800\times10^6\cr
B_4       |-1.13\times10^3  | -7.80\times10^3       | -2.65\times10^4
| -6.64\times10^4  |-1.09\times10^6\cr
C_4       | 612    | 2.87\times10^3     | 7.35\times10^3    | 1.48\times10^4
|1.23\times10^5       \cr
D_4       | -75.2 | -241   | -462   | -742  | -3060      \cr
E_4| -13.0| -13.3 | -15.8| -27.9 | -50.6
\endtable
\bigskip
\leftskip = 25 pt\rightskip = 10pt
{\it \noindent
Table~VIII: WAPAP's for the five-loop NSVZ $\beta$-function,
with $E_4$ not input. The values of $N_F^{max}$ used range between
4 and 25.}
\out}
\bigskip

We see that the
WAPAP's are in general in good agreement with the exact result for
$A_4$ in the NSVZ scheme, at the 10\% level.
Although encouraging, these results are not quite as compelling as the
ones for the DRED scheme. This is at first sight surprising,
given the form of \russa, which appears at first sight to be
close to the rational function form of the PA's. However,
as mentioned in the Introduction, perhaps minimal subtraction
schemes are more amenable to Pad\'e techniques.
The anomalously poor result for $A_4$ in Table~VII is caused by the fact
that
the error $\delta_4$ is close to $-1$ in this case, for the
$N_F^{max}$ values corresponding to the determination of $D_4$.
The reason the result for $D_4$ is not also anomalously large is that
the two values from which the weighted average is taken are both numerically
large but with opposite signs. Thus we cannot rely on either the $A_4$ or the
$D_4$ prediction for $N_C = 5$. With this exception,
$A_4$ comes out reasonably close to the exact result.
This means, of course that the predictions for $B_4\cdots D_4$ will not
change much if we input $A_4$ as well as $E_4$.

Analogously to the five-loop QCD case discussed in the previous
section, we take the differences between the entries in
Tables~VII and VIII as lower limits on the possible uncertainties
in our five-loop NSVZ predictions.

\newsec{The Quark Mass Anomalous Dimension in QCD}
We now consider the quark mass anomalous dimension $\gamma$ in QCD,
defined as
\eqn\gama{
\gamma={d \ln m_q\over{d\ln \mu^2}}=-\gamma_0a-\gamma_1a^2-\gamma_2a^3
-\gamma_3a^4-\gamma_4a^5+O(a^6),}
where $a={\alpha_s/{\pi}}$. The 4-loop coefficient $\gamma_3$ was recently
computed in~\ref
\larinb{J.A.M.~Vermaseren, S.A.~Larin and T.~van Ritbergen,
\plb405 (1997) 327.}
\ref\chetyrkin{K.G. Chetyrkin,
\plb404 (1997) 161.}
and the full exact results for the coefficients $\gamma_n$
for $n=0,1,2,3$ are given by
\eqn\gamb{\eqalign{
\gamma_{0} & = \frak{1}{4}\left[ 3 C_F \right]\cr
 \gamma_{1} & = \frak{1}{16} \left[
 \frak{3}{2}C_F^2+\frak{97}{6} C_F C_A -\frak{10}{3} C_F T_F N_F
  \right] \cr
 \gamma_{2} & = \frak{1}{64} \Bigl[
  \frak{129}{2} C_F^3 - \frak{129}{4}C_F^2 C_A
 + \frak{11413}{108}C_F C_A^2  \cr &
 +C_F^2 T_F N_F (-46+48\zeta_3)
+C_F C_A T_F N_F \left( -\frak{556}{27}-48\zeta_3 \right)
- \frak{140}{27} C_F T_F^2 N_F^2  \Bigr]
   \cr
\gamma_{3} & =  \frak{1}{256} \Bigl[
  C_F^4 \left(-\frak{1261}{8} - 336\zeta_3 \right)
   + C_F^3 C_A \left( \frak{ 15349}{12} + 316 \zeta_3 \right)
 \cr&
   + C_F^2 C_A^2 \left(-\frak{ 34045}{36} - 152 \zeta_3 + 440\zeta_5 \right)
   + C_F C_A^3 \left( \frak{70055}{72} + \frak{1418}{9} \zeta_3
                      - 440 \zeta_5 \right)
 \cr&
  + C_F^3 T_F N_F \left( -\frak{280}{3} + 552 \zeta_3 - 480 \zeta_5 \right)
  + C_F^2 C_A T_F N_F \left(- \frak{8819}{27} + 368 \zeta_3
                            - 264 \zeta_4 + 80 \zeta_5 \right)
 \cr&
  + C_F C_A^2 T_F N_F \left(- \frak{65459}{162}
                  - \frak{2684}{3} \zeta_3 + 264 \zeta_4
                    + 400 \zeta_5 \right)
 \cr&
  + C_F^2 T_F^2 N_F^2 \left( \frak{304}{27} - 160 \zeta_3
                            + 96 \zeta_4 \right)
 \cr&
  + C_F C_A T_F^2 N_F^2 \left( \frak{1342}{81}
                             + 160 \zeta_3 - 96 \zeta_4 \right)
  + C_F T_F^3 N_F^3 \left(- \frak{664}{81} + \frak{128}{9} \zeta_3 \right)
 \cr&
  + \frakk{d_F^{a b c d}d_A^{a b c d}}{d_Q}
            \left(- 32 + 240 \zeta_3  \right)
  +  N_F \frakk{d_F^{a b c d}d_F^{a b c d}}{d_Q}
            \left( 64 - 480 \zeta_3  \right) \Bigr].
\cr}}
where for $SU(N_C)$ the quadratic and quartic Casimirs are as defined in
\qcdc, \qcdd, and $T_F=\frak{1}{2}$ as before.
In addition, $d_Q$ is the dimension of
the quark representation, so that $d_Q=N_C$ for $SU(N_C)$, and we have
$\zeta_4\equiv\zeta(4)= 1.0823232\cdots$ and $\zeta_5\equiv\zeta(5)
= 1.0369277\cdots $. For $N_C=3$, we
have
\eqn\gambb{\eqalign{
 \gamma_0 &= 1   \cr
\gamma_1 & =  \frak{1}{16}\left[ \frak{202}{3}
                 - \frak{20}{9} N_F \right]\cr
 \gamma_2 & =  \frak{1}{64} \left[1249+\left( - \frak{2216}{27}
          - \frak{160}{3}\zeta_3 \right)N_F
               - \frak{140}{81} N_F^2 \right]\cr
 \gamma_3 & =  \frak{1}{256} \Bigl[
       \frak{4603055}{162} + \frak{135680}{27}\zeta_3 - 8800\zeta_5
 +\left(- \frak{91723}{27} - \frak{34192}{9}\zeta_3
    + 880\zeta_4 + \frak{18400}{9}\zeta_5 \right) N_F\cr
 &+\left( \frak{5242}{243} + \frak{800}{9}\zeta_3
    - \frak{160}{3}\zeta_4 \right) N_F^2
 +\left(- \frak{332}{243} + \frak{64}{27}\zeta_3 \right) N_F^3 \Bigr],
\cr}}
which have the numerical values
\eqn\gamc{\eqalign{
 \gamma_0 & = 1 \cr
 \gamma_1 & \approx 4.20833 - 0.138889 N_F  \cr
 \gamma_2 & \approx 19.5156 - 2.28412 N_F - 0.0270062 N_F^2  \cr
 \gamma_3 & \approx 98.9434 -19.1075 N_F+ 0.276163 N_F^2
                       +0.00579322 N_F^3.\cr}}
Omitting the quartic Casimir contributions, one obtains
\eqn\gamd{
\gamma_3=96.4386-18.8292N_F+0.276163N_F^2+0.00579322N_F^3.}
and we shall now compare \gamc\ and \gamd\ with APAP's.

It transpires that the WAPAP procedure does not work so well here. The most
accurate results for both $B_3^{\gamma}$ and $C_3^{\gamma}$ are obtained for
small $N_F^{max}$. This is reasonably consistent with the WAPAP behaviour
in the $C_3^{\gamma}$ case:
here, the weight difference
$C_3^{\gamma W}-A_3^{\gamma W}$ never changes
sign, but is smallest at $N_F^{max}=2$ on the edge of the range. However, the
WAPAP criterion for $B_3^{\gamma}$ leads to values of $N_F^{max}$ which start
at 9  for $N_C=2$ and increase with $N_C$.
Nevertheless, as in the previous sections, it seems sensible  to match at
negative $N_F$, and spectacular results are
obtained if we simply take $N_F^{max}=4$  (with $-N_F^{max} < N_F < 0$)
throughout, as
can be seen from Table~IX, where numerical
predictions for the coefficients in the parametrization
\eqn\game{
\gamma_3=A^{\gamma}_3+B^{\gamma}_3N_F+C^{\gamma}_3N_F^2+D^{\gamma}_3N_F^3,}
are given both without (w/o Q) and with (w. Q) quartic Casimir
contributions.
It should be noted that we have used as input the
exact result for $D_3^{\gamma}$, which is contained in~\gambb.

\bigskip
\vbox{
\noncenteredtables
\tablewidth=30.2em
\line{\kern6.58em
\begintable
\multispan{5} \strut $N_C$ \cr
\kern 0em 2 \kern0.2em|\kern0em 3\kern0.0em|\kern0em 4\kern0em| \kern-1.2em
5\kern-1.2em| \kern-0.2em 20 \kern0em
\endtable
}
\centeredtables
\tablewidth=-\maxdimen
\thicksize=0.0pt 
\begintable
\bb &\multispan{6} \mystrut \quad $A_3^{\ga}$ \hfill                 & \bb \cr
\bb |\hb{APAP}      | 16.1   | 97.9   | 328   | 822  | 2.18 \times
10^5 | \bb \cr
\bb |\hb{w/o Q}     | 15.4   | 96.4   | 327   | 825  | 2.23 \times
10^5 | \bb \cr
\bb |\hb{w. Q}      | 16.0   | 98.9   | 334   | 840  | 2.26 \times
10^5 | \bb \cr
\bb &\multispan{6} \mystrut \quad $B_3^{\ga}$ \hfill                 & \bb \cr
\bb |\hb{APAP}      | -5.14   |-20.0   |-49.3   |-98.0  |-6.39 \times 10^3
| \bb \cr
\bb |\hb{w/o Q}     | -4.70   |-18.8   |-47.1   |-94.2  |-6.27 \times 10^3
| \bb \cr
\bb |\hb{w. Q}      | -4.77   |-19.1   |-48.0   |-96.2  |-6.43 \times 10^3
| \bb \cr
\bb &\multispan{6} \mystrut \quad $C_3^{\ga}$ \hfill                 & \bb \cr
\bb |\hb{APAP}      | 0.065   | 0.224   | 0.478   | 0.828  | 17.5   |
\bb \cr
\bb |\hb{exact}     | 0.111   | 0.276   | 0.504   | 0.796  | 13.0   | \bb
\cr
\bb & \multispan{6} \mystrut \quad $D_3^{\ga}$ \hfill                & \bb \cr
\bb | \hb{input}    | 3.26 \times 10^{-3} | 5.79 \times 10^{-3} | 8.15
\times 10^{-3} | 0.0104 | 0.0433  | \bb \cr
\bb & \multispan{6} & \bb
\endtable
\thicksize=0.7pt 
\bigskip
\in
{\it \noindent Table~IX: Four-loop quark mass anomalous dimension in QCD:
APAP's  for fixed $N_F^{max} = 4$ are compared with the exact values both
without (w/o Q) and with (w. Q) the quartic Casimir terms.}
\bigskip
\out
} 
\vskip3ex

It can be seen that in
all cases the APAP estimate is quite accurate over
a wide range of $N_C$. In most cases, the APAP estimate
is closer to the exact result without the quartic Casimir contribution
(w/o Q),
but in any case the quartic Casimir contribution to $\gamma_3$ is smaller
than in the case of the QCD $\beta$ function.

We now go on to discuss the five-loop APAP estimate of $\gamma$. We
parametrize
the five-loop quark mass anomalous dimension $\gamma_4$ in the form
\eqn\gamf{
\gamma_4=A^{\gamma}_4+B^{\gamma}_4N_F+C^{\gamma}_4N_F^2+
D^{\gamma}_4N_F^3+E^{\gamma}_4N_F^4,}
where the value of $E^{\gamma}_4$ can be derived
from~\ref\pmpp{A. Palanques-Mestre  and P.~Pascual, \cmp 95 (1984) 277.}:
\eqn\gamg{
E^{\gamma}_4 = C_F T_F^4(-65/5184 -5\zeta(3)/324 + \pi^4/3240)\ .}
We use the full $\gamma_3$ as input, including the quartic Casimir
contribution. As we argued in the case of the QCD $\beta$ function, we expect
our five-loop estimate to include the effects of contributions involving such
quartic Casimir terms, but not the effect of new Casimir terms making a
first appearance. Once again we choose $N_F^{max}=4$ to derive the results
shown in Table~X.
\vskip3em
\vbox{
\begintable
N_C       |  2         |     3      |       4   |   5       | 20         \cr
A_4 \wQ | 56.0      |  530     | 2.41 \times 10^3   | 7.63 \times
10^3
| 8.37\times10^6  \cr
A_4 \woQ | 50.5      |  493     | 2.27 \times 10^3  | 7.22 \times
10^3
| 7.97\times10^6  \cr
B_4 \wQ |-23.3      | -143     | -483    | -1.22 \times 10^3
| -3.33\times10^5   \cr
B_4 \woQ  |-21.7      | -135     | -457    | -1.15 \times 10^3
| -3.12\times10^5   \cr
C_4 \wQ   | 1.70      | 6.67      | 16.8     | 33.7     |2.29 \times
10^3 \cr
C_4 \woQ| 1.64      | 6.44      | 16.0     | 32.0     |2.14 \times
10^3 \cr
D_4 \wQ  | 8.12\times10^{-3}   | 0.037    | 0.0891   | 0.165    | 4.31      \cr
D_4 \woQ | 8.88\times10^{-3}   | 0.037    | 0.0831   | 0.148    | 3.48      \cr
E_4(\hb{input)}| -4.80\times10^{-5} | -8.54\times10^{-5} | -1.2\times10^{-4}
| -1.54\times10^{-4} | -6.39\times10^{-4}
\endtable
\bigskip
\leftskip = 25 pt\rightskip = 10pt
{\it \noindent
Table~X: APAP's for the five-loop quark mass anomalous
dimension in QCD, calculated with and without the four-loop quartic
Casimir terms.}
\out}
\bigskip

\newsec{Abelian Gauge Theories}

All of the previous sections have dealt with APAP predictions for
{\it non-Abelian\/} theories. It is natural to ask whether similarly
accurate results can be obtained for the Abelian case.
We address this question  in this section,
choosing as our example the fermion mass
anomalous dimension with $N_F$ charged fermions,
where good results were found in the non-Abelian case,
as we saw in the previous section. A supplementary
reason for choosing this example is that the {\it
next-to-leading\/}-$N_F$ result is available,
as well as the leading one.

The results for $\gamma_1\cdots\gamma_3$ in the abelian case follow from
Eq.~\gamb\ by setting
\eqn\caina{
C_F=T_F=1, C_A =0, \frakk{d_F^{a b c d}d_A^{a b c d}}{d_Q}=0,
\frakk{d_F^{a b c d}d_F^{a b c d}}{d_Q}=1,}
so that:
\eqna\cainb$$\eqalignno{
  \gamma_0 & = 0.75 & \cainb a\cr
 \gamma_1 & \approx 0.09375 - 0.2083 N_F & \cainb b \cr
 \gamma_2 & \approx 1.0078 +0.18279 N_F - 0.08102 N_F^2 & \cainb c \cr
 \gamma_3 & \approx -2.1934 -1.7207 N_F-0.30143 N_F^2
                       +0.03476 N_F^3.& \cainb d\cr}$$
Omitting the quartic Casimir term, we would instead have
\eqn\cainc{\gamma_3  \approx -2.1934 + 0.2831 N_F-0.30143 N_F^2
                       +0.03476 N_F^3.}
We can see at once that the miraculous success of the previous APAP
prediction for  $\ga_3$  will not be reproduced here. For $N_F=0$,
the {\it sign\/} of $\ga_3$  differs from the sign of $\ga_2^2/\ga_1$.
Moreover, $\ga_1$ has a zero, and hence $\ga_3^{\rm APAP}$ has a pole,
for $N_F\approx 0.45$. Hence, we cannot hope to reproduce $\ga_3$ for
small
values of $|N_F|$. For large $|N_F|$ the sign of $\ga_3$ is still wrong,
so the method fails in this region also.

One easily verifies that this pessimism is confirmed by the results,
and things do not improve at five loops.
Then, as well as $E_4^{\ga}$ as given in~\gamg\ , it is possible to
derive from ~\ref\jag{J.A.~Gracey,
\plb317 (1993) 415.}\ the result for $D_4$:
\eqn\caine{
D_4^{\ga} = \frakk{11}{96}\zeta_3+\frakk{1}{6}\zeta_{5}-\frakk{\pi^4}{288}
+\frakk{4483}{41472}\approx0.0804.}
We notice now, however, that $\gamma_2$ has zeros,
and hence $\gamma_3^{\rm APAP}$ has poles, for $N_F=-2.6$ and $N_F= 4.8$.
Consequently,
we may expect that the results will be rather sensitive to the range of
$N_F$, if we match
in a region including the origin. On the other hand, for large $N_F$ we
have $\ga_3^2/\ga_2 \approx -0.014N_F^4$, whereas $E_4^{\ga}\approx
-0.001$,
so we also cannot expect good results at increasing $|N_F|$.

We leave it to the reader to convince her(him)self that we cannot expect to
extract reliable predictions for $A_4\cdots C_4$.
We also record that
the QED and SQED gauge $\beta$-functions yield
similarly unattractive results.
Evidently, Abelian theories are less amenable to the APAP
approach, for some unknown reason.

\newsec{Conclusions}
We have presented results obtained from our APAP method for the
four-loop and five-loop QCD $\beta$-function coefficients,
for the five-loop SQCD $\beta$-function coefficients, and for the
four- and five-loop quark mass anomalous dimensions in QCD.
Particularly in the case of the QCD $\beta$-function,
and to some extent also for SQCD, particularly in the DRED scheme,
a modified procedure for extracting the predictions for the
various coefficients  of powers of $N_F$ (WAPAP) gave improved results.
In general, the four-loop results
agree very well with the known results, giving us
confidence in our predictions of the five-loop terms.

Our four-loop QCD $\beta$-function
predictions~\EKS\ were confirmed very rapidly by
an exact calculation~\larin. Unfortunately,
in view of the current limitations on the technology of exact
perturbative calculations in QCD and SQCD, it may be some time before
our five-loop predictions can also be tested directly.
It would therefore be interesting to find alternative
techniques that could be confronted or combined with APAP's.
One possible complementary technique may be that of the large-$N_F$
expansion.
Unfortunately, it is the leading term in $N_F$ which is least well
determined by the APAP approach, which is related to the
poor results obtained in the Abelian case.
It would be very interesting if the large-$N_F$
methods could be extended
to next-to-leading terms in this expansion for the non-Abelian case,
in which case more
comparisons and cross-checks could be made.
\vskip 0.2in
\centerline{\bf Acknowledgements}
\vskip 0.2in
This work was supported by the US Department of Energy under contract
DE-AC03-765F00515 and grant no. DE-FG05-84ER40215.
The research of M.K.  was supported in part by the Israel
Science Foundation administered by the Israel Academy of Sciences and
Humanities, and by a Grant from the G.I.F., the German-Israeli
Foundation for Scientific Research and Development.
I.J. and D.R.T.J. thank
John Gracey for help with his large-$N_F$ results.

\footatend\vfill\supereject\immediate\closeout\rfile\writestoppt
\baselineskip=14pt\centerline{{\bf References}}\bigskip{\frenchspacing%
\parindent=20pt\escapechar=` \input refs.tmp\vfill\eject}\nonfrenchspacing
\bye